\documentclass{emulateapj}

\shorttitle{X-ray Emission from the M87 Jet}
\shortauthors{Harris et al.}

\begin{document}


\title{Flaring X-ray Emission from HST-1, a Knot in the M87 Jet}


\author{D. E. Harris\altaffilmark{1}, J. A. Biretta\altaffilmark{2},
W. Junor\altaffilmark{3}, E. S. Perlman\altaffilmark{4},
W. B. Sparks\altaffilmark{2}, and A.~S.~Wilson\altaffilmark{5, 6}}

\altaffiltext{1}{Smithsonian Astrophysical Observatory, 60 Garden St.,
Cambridge, MA 02138, harris@cfa.harvard.edu}

\altaffiltext{2}{Space Telescope Science Institute, Baltimore, MD 21218, biretta@stsci.edu}

\altaffiltext{3}{Space \& Atmospheric Sciences, Los Alamos National
Laboratory,Los Alamos, NM 87545, bjunor@lanl.gov}

\altaffiltext{4}{Joint Ctr. for Astrophysics, University of Maryland, Baltimore
County, Baltimore, MD  21250, perlman@jca.umbc.edu}

\altaffiltext{5}{Astronomy Department, University of Maryland, College Park, 
MD 20742, wilson@astro.umd.edu}

\altaffiltext{6}{Adjunct Astronomer, Space Telescope Science Institute,
3700 San Martin Drive, Baltimore, MD 21218}




\begin{abstract}
We present Chandra X-ray monitoring of the M87 jet in 2002, which
shows that the intensity of HST-1, an optical knot 0.8$''$ from the core,
increased by a factor of two in 116 days and a factor of four in 2
yrs.  There was also a significant flux decrease over two months, with
suggestive evidence for a softening of the spectrum.  From this
variability behavior, we argue that the bulk of the X-ray emission of
HST-1 comes from synchrotron emission.  None of the other conceivable
emission processes can match the range of observed characteristics.
By estimating synchrotron model parameters for various bulk relativistic
velocities, we demonstrate that a model with a Doppler factor $\delta$
in the range 2 to 5 fits our preliminary estimates of light travel
time and synchrotron loss timescales.

\end{abstract}




\section{Introduction}

With the advent of Chandra X-ray observations of radio jets, evidence
supporting the synchrotron process for the origin of the X-ray
emission from the jets of low luminosity radio galaxies has grown.
One line of evidence is the similarity of the X-ray morphology to the
radio and optical structures which are most likely synchrotron
emission because of the detection of linear polarization. Another
argument is that the values of the spectral index of power law fits to
the X-ray spectra are generally steeper than the radio spectra whereas
inverse Compton (IC) models predict
$\alpha_x\leq\alpha_r$\footnote{The spectral index, $\alpha$, is
defined by flux density, S~$\propto~\nu^{-\alpha}$.} (e.g. Hardcastle
et al. 2002; Harris \& Krawczynski, 2002).

In this paper we exploit another fundamental difference between IC and
synchrotron emission: the variability time scale.  Although the
injection of new relativistic electrons can produce the same rise time
for both processes, the characteristic loss timescales are vastly
different.  All IC models require relatively low energy relativistic
electrons.  For synchrotron self Compton (SSC) models, Lorentz energy
factors, $\gamma$, are normally comparable to the values of a few
thousand which are responsible for the observed radio emission.
For IC scattering between the relativistic electrons and photons
of the cosmic microwave background (IC/CMB), $\gamma\leq$1000
(Harris \& Krawczynski 2002).  Synchrotron models however require
$\gamma\approx10^7$ with corresponding half-lives of order a year for
the sorts of magnetic field strengths commonly estimated for
features in radio jets.

Our data show strong intensity variability from the unresolved core
and HST-1.  We use the flux doubling time to derive an upper limit to
the source size and the decay time to constrain the timescale for
E$^2$ losses, and in turn, $\delta$.  We take the distance to M87 to
be 16 Mpc so that one arcsec~=~77pc.

\section{The X-ray monitoring observations\label{sec:data}}

\begin{figure*}
\epsscale{0.8}
\plotone{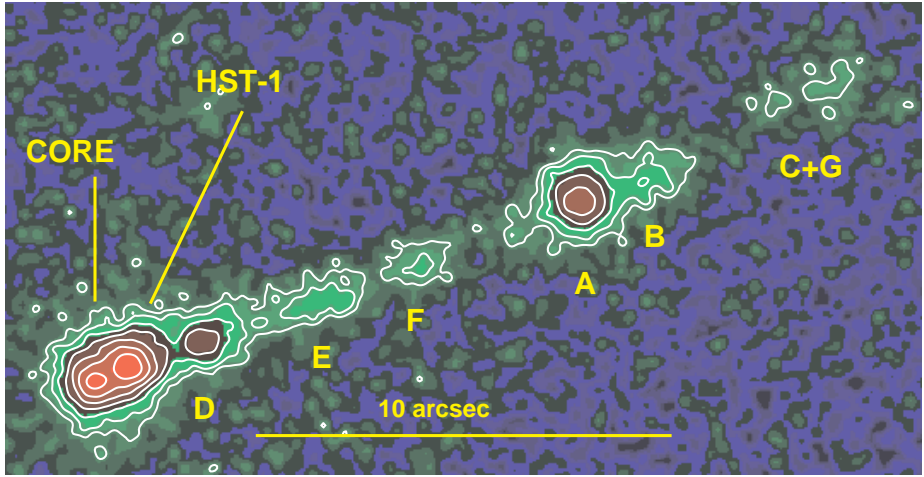}
\caption{Chandra X-ray image of the M87 jet constructed by averaging
the data of Wilson \& Yang (2002) taken in 2000 July, together with
our 5 observations.  The contours increase by factors of two in
brightness, with the lowest contour level being
1$\times$10$^{-16}$~erg~cm$^{-2}$~s$^{-1}$~per pixel in the 0.2 to 6
keV band.  Jet features are labelled according to the standard
convention and a 10$''$ scalebar is shown.  ACIS pixel randomization
has been removed; the data were regridded to a pixel size of
0.0492$^{\prime\prime}$; and a Gaussian smoothing function of
FWHM=0.25$^{\prime\prime}$ was applied.}
\label{fig:sum6}
\end{figure*}

In 2002 we obtained five observations of 5ks each on Jan16, Feb12,
Mar30, Jun08, and Jul24.  To minimize pileup, we used a readout time
of 0.4s for the standard 1/8th subarray on the ACIS-S3 chip (PUG
2001).  Data reduction followed the
'threads'\footnote{http://asc.harvard.edu/ciao/threads/} (CIAO2.2).
To recover the inherent resolution of Chandra, we removed pixel
randomization and rebinned the data to 0.1 native ACIS pixels.  For 3
energy bands (soft, S=0.2-0.75keV, medium, M=0.75-2keV, and hard,
H=2-6keV) flux maps were obtained by standard procedures; a correction
was applied for degradation of the quantum efficiency by measurement
of fixed regions of hot gas emission near the jet.  A wide band image
was made by adding the 3 energy bands and fig.~\ref{fig:sum6}
shows the result of averaging our fluxmaps.

\begin{figure*}
\epsscale{0.9}
\plottwo{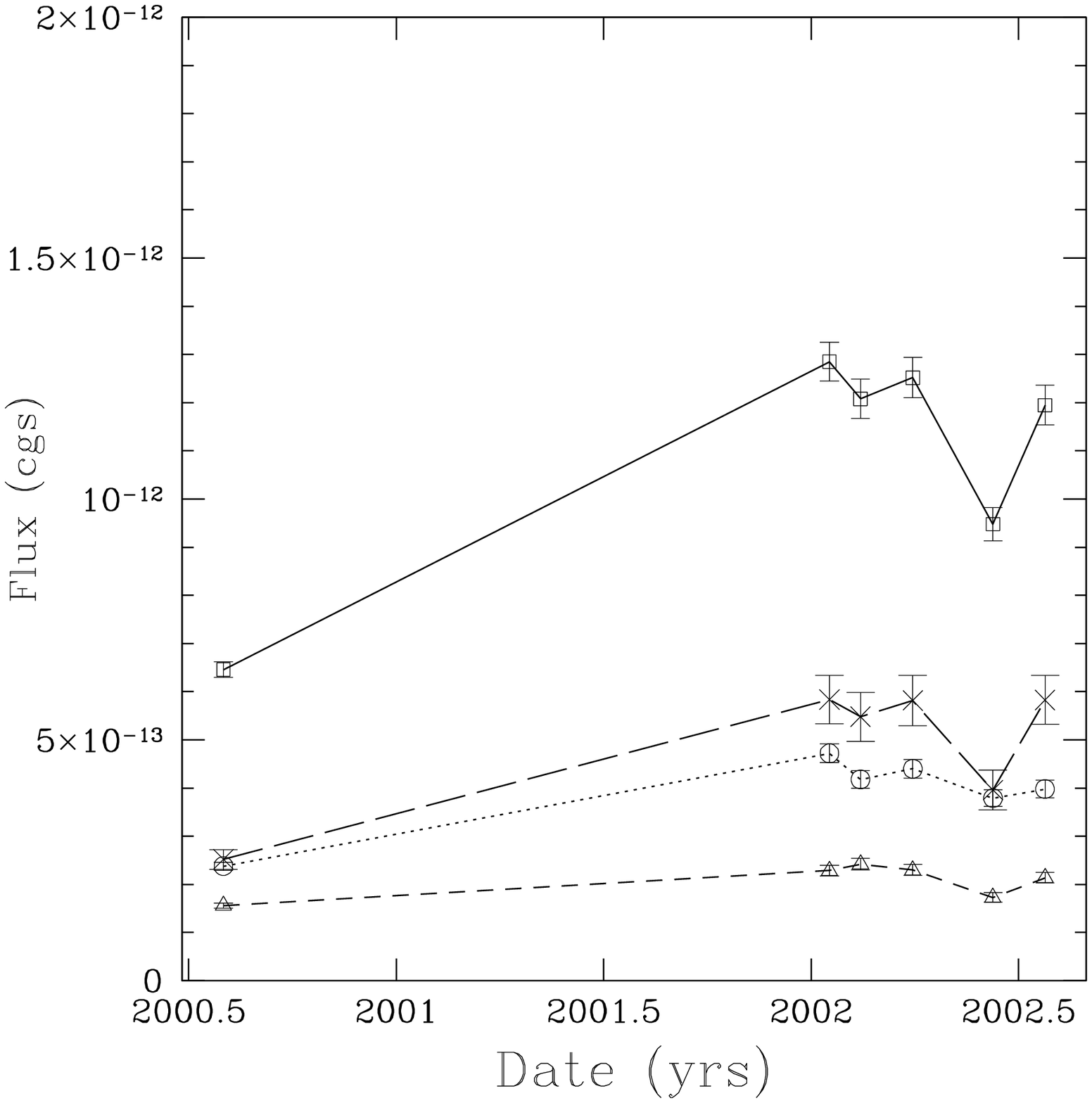}{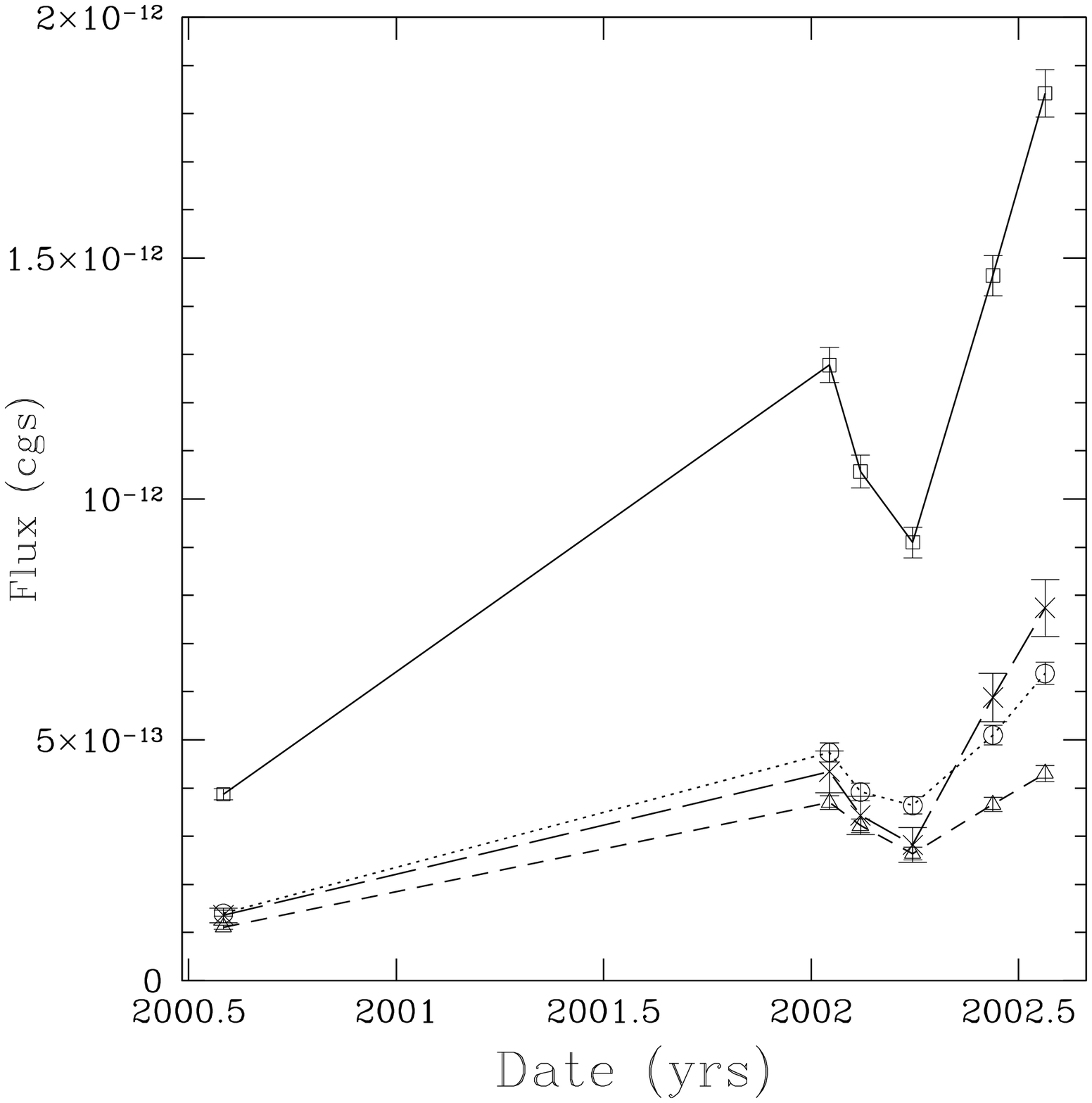}
\caption{Lightcurves for the core(left panel) and HST-1 (right panel).
The curves are for different energy bands: total (solid line), hard
(broken), medium (dotted), and soft (dashed).  The data at the
left edge come from the archival 2000 July observation (Wilson and Yang
2002) and those to the right from our 2002 monitoring.}
\label{fig:lc}
\end{figure*}

Our monitoring shows striking variability in both HST-1 and the
unresolved core (which could well come from jet segments closer than
0.4$^{\prime\prime}$ to the nucleus).  The other knots show weak or no
variability and will be analyzed in a future paper.  Contour diagrams
of the core and HST-1 for each observation can be found in Harris
(2003a); here we show lightcurves for the
core and HST-1 (fig.~\ref{fig:lc}).  The
photometry was obtained by taking small circles with radii
0.44$^{\prime\prime}$ and subtracting the background from rectangles
above and below the jet (with a total area of 13.6~arcsec$^2$).

The largest increase occurs for HST-1 between Mar and Jul
with over a factor of two brightening in 116d\footnote{A
brightening was also seen in our HST monitoring between 2001 July and
2002 July.  Those data will be analyzed in a separate paper (Biretta et
al. 2003, in preparation)}.  This implies that the emitting volume of
the variable component has a characteristic size $\leq$~the light
travel time\footnote{Although see Protheroe (2002) for caveats.} which
is 0.1pc for a stationary source.  The brightening in the core between
our last two observations showed a similar rate of increase with a
20\% gain in 46 days.

The relative increases in the three energy bands demonstrate a
hardening of the spectrum for HST-1 between March and July 2002
(compare the rates of increase in each band in fig.~\ref{fig:lc}).
The intensity increase was a factor of 2.7$\pm$0.3 for the hard,
1.7$\pm$0.1 for the medium, and 1.6$\pm$0.1 for the soft band.
Although we defer spectral analysis to a later paper, this hardening
produced a change in the X-ray spectral index from
$\alpha_x\approx1.35\pm0.1$ (prior to 2002 May) to
$\alpha_x=1.05\pm0.07$ (for 2002 July).

The 'cleanest' drop occurs for HST-1 between the 2002 January and
February observations.  The intensity ratios (Feb over Jan) for these
27 days are: 0.79$\pm$0.09 (H), 0.83$\pm$0.04 (M), and 0.87$\pm$0.05
(S).  While these values are the same within the errors,
there is a progression consistent with expectations for E$^2$ losses
(but contrary to that of increased absorption): the hard band is
dropping faster than the soft band.  A similar effect is seen in the
core decay between Mar and Jun: the hard band dropped 32$\pm$7\% while the
medium band dropped 14$\pm$5\%.

All of these timescales are subject to considerable error.
Although the statistical errors for the flux values are
less than 10\%, there will always be some flux present which is not
part of the variable component and our sampling interval is not
small enough to determine shorter variations.

Variability of a synchrotron source can result from a number of
effects such as changes in any of the absorption along the line of
sight, the magnetic field strength, B, the number of particles
contributing to the radiation in a given band, or the beaming factor
(Protheroe 2002).

Adiabatic compression or expansion is particularly effective in
changing the radiated power since the electrons gain or lose energy
and at the same time, B changes to augment the effect.  Given a power
law electron energy distribution N(E)$\propto$E$^{-p}$, there should
be no change in p.  On the other hand, if the effects of a high energy
cutoff extend down to the energy range of interest, a compression
would cause an increase of intensity accompanied by a spectral
hardening since the emission from a fixed observing band would come
from lower energy electrons and the spectral curvature would be
shifted to higher energies.  However, for the remainder of this paper
we focus on an alternative synchrotron model based on changes to the
number of radiating particles.

For this model, increased intensity arises from the injection of new
radiating particles with a spectrum flatter than that which existed
previously.  The spectral softening which accompanies flux decline
comes from normal $E^2$ losses.  The chief difference in these models
is that the compression/expansion model requires that N(E) departs
from a power law.  Our data allow spectral curvature but the
statistics are not good enough to demand curvature.

\section{The Spectral Energy Distribution of HST-1}

In order to derive synchrotron model parameters, we need to know
the overall spectral distribution of the emitted energy.
Both the radio and optical morphologies of the inner jet are shown in
fig. 3 of Perlman et al. (1999) with 0.2$^{\prime\prime}$ resolution.
At higher resolution (fig. 1 of Biretta, Zhou, \& Owen 1995) it can
be seen that there are a series of several knots between
0.8$^{\prime\prime}$ and 1.2$^{\prime\prime}$ from the core.  HST data
demonstrate that the brightest upstream knot is slowly moving but the
downstream knots are moving with an apparent speed of 6c (Biretta,
Sparks, \& Macchetto 1999).  Our feature 'HST-1'
(fig.~\ref{fig:sum6}) corresponds to the upstream, optically brighter
part of this segment of the jet.  We take the radio flux density
corresponding to the X-ray knot to be 3.8 mJy, as measured on the high
resolution 15 GHz map.  From a new VLA observation,\footnote{We
obtained a short observation in 2002 October as a result of an ad hoc
proposal.  The VLA was in C array with a beamsize of
0.5$^{\prime\prime}$.}  we derive an upper limit of 2 mJy at 43 GHz.

\begin{figure}
\vspace{-1.0in}
\epsscale{1.2}
\plotone{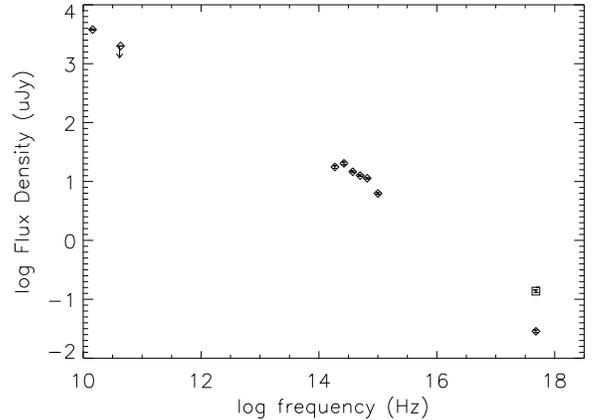}
\caption{The spectrum of HST-1.  Optical flux
densities come from HST data (1998) of Perlman et al. (2001), but with
smaller apertures to correspond to just the first, brighter, part of
the complex.  The X-ray flux densities come from the observed fluxes
(i.e. no correction for absorption has been applied) assuming
$\alpha_x$=1.  Values are shown for the faintest (2000 July) and
brightest (2002 July) intensities.  The actual flux densities may be
somewhat larger since we used a small aperture for photometry to
isolate the variable fraction of the total intensity.}
\label{fig:spec}
\end{figure}

The optical flux densities come from the detailed spectral analysis at
0.15$^{\prime\prime}$ resolution based on data obtained in 1998 February by
Perlman et al. (2001).  The spectrum is shown in fig.~\ref{fig:spec},
and for the purposes of sec.~\ref{sec:sync},
we approximate these data with a single power law between 10$^9$ and
10$^{18}$~Hz with $\alpha$=0.68 and amplitude determined by
S(2keV)=100nJy, parameters which describe the observables to within a
factor of two.

\section{A 'modest beaming' Synchrotron Model for HST-1\label{sec:sync}}

Previous analyses have demonstrated that the X-ray jet emission of M87
is not thermal emission (Wilson \& Yang 2002; Marshall et al. 2002;
Biretta, Stern, \& Harris 1991); is not SSC emission (ibid.); and is
not IC/CMB emission because of the large values of $\Gamma$ and small
angles to the line of sight required (Harris \& Krawczynski 2002).
The X-ray spectra of the knots are much steeper than in the radio or
optical and strongly favor synchrotron radiation (Wilson \& Yang
2002), confirming the conclusion of Biretta, Stern, \& Harris (1991).
Our variability data is fully consistent with X-ray synchrotron
emission and it remains only to demonstrate that plausible physical
parameters can be found with a simple model which posits that the jet
has a relativistic bulk velocity, as implied by the one-sided
emission at all frequencies.  In this section, we do that for HST-1.


Using the time required for the flux to double provides an upper limit
to the source size.  To use the observed decline, we take the observed
variables back to the jet frame, calculate the properties of the
synchrotron source assuming equipartition conditions (Pacholczyk
1970), and then evaluate the halflife of the electrons responsible for
the X-ray emission, both in the jet frame and as observed at the
Earth.  We also compare energy densities in the magnetic field and in
the synchrotron photons.  We performed these calculations for 8 values
of $\delta$ between 1 and 16.  For converting various quantities to
the jet frame and back, we used expressions from the appendix of
Harris and Krawczynski (2002).  The results for representative values
of $\delta$ are shown in Table~\ref{tab:results}.


\begin{deluxetable*}{lcccccc}
\tablecaption{Beamed Synchrotron Parameters for HST-1\label{tab:results}}
\tablewidth{0pt}
\tablehead{
\colhead{Parameter} & \colhead{$\delta$=1}   & \colhead{$\delta$=2}   &
\colhead{$\delta$=4} & \colhead{$\delta$=6}  &
\colhead{$\delta$=8} & \colhead{$\delta$=16}\\
}
\startdata
radius (arcsec) & 0.0012 & 0.0024 & 0.0048 & 0.0072 & 0.0096 & 0.0192\\
L$_s^{\prime}$ (erg~s$^{-1}$) & 6.5E40 & 4.1E39 & 2.5E38 & 5.3E37 & 1.6E37 & 1.0E36\\
log E$^{\prime}_{tot}$ (ergs) & 48.55 & 48.40 & 48.17 & 48.04 & 47.96 & 47.75\\
$\gamma^{\prime}_1$ & 135 & 181 & 244 & 274 & 328 & 443\\
$\gamma^{\prime}_2$ & 4.2E6 & 5.7E6 & 7.7E6 & 8.7E6 & 1.0E7 & 1.4E7\\
B$^{\prime}$ (mG) & 13 & 3.7 & 1.0 & 0.51 & 0.28 & 0.077\\
u$^{\prime}$(B) (erg~cm$^{-3}$) & 6.9E-6 & 5.4E-7 & 4.0E-8 & 1.0E-8 & 3.7E-9 & 2.4E-10\\
u$^{\prime}$(sync) (erg~cm$^{-3}$) & 7.5E-6 & 1.2E-7 & 1.8E-9 &
1.9E-10 & 2.9E-11 & 4.5E-13\\
$\tau_\frac{1}{2}^{\prime}$ (yr) & 0.011 & 0.126 & 1.3 & 4.3 & 12.1 & 81.9\\
$\tau_\frac{1}{2}$(Earth) (yr) & 0.011 & 0.063 & 0.32 & 0.72 & 1.5 & 5.1\\

\enddata




\tablecomments{The radius is the angular size corresponding to the
light travel time in the jet frame.  Jet frame parameters are primed.
L$_s^{\prime}$ is the synchrotron luminosity integrated over the band
which is observed at the Earth between E9 and E18Hz.  E$_{tot}$ is the
total energy in particles and fields required to produce
L$_s^{\prime}$.  $\gamma_{1,2}$ are the lower and upper bounds on the
electron energy distribution.  The u's are the energy densities in the
magnetic field and the synchrotron photons.  $\tau_\frac{1}{2}$ is the
halflife of electrons producing the highest energy photons.}


\end{deluxetable*}

We have made model specific estimates of how to relate an observed
decline of 21\% (in the hard band in 27 days) to the effects of E$^2$
losses on the electron distribution (Harris
2003b).  We find consistency for $\delta$ in the range 2 to 5.
However, the main conclusion is that regardless of the precise model
and the actual values found for the rate of decline, the observed
timescales for decay are fully consistent with the beaming parameters
for a jet angle of order 10$^{\circ}$ to 17$^{\circ}$ (to the line of
sight) suggested by the HST proper motion studies of Biretta et
al. (1999) in their Table 3.  Larger beaming factors imply weaker
fields and longer halflives, and vice-versa.  A recent HST observation
shows the varying region to be unresolved in the optical/UV,
independently placing an upper limit of $0.02''$ (1.5pc) on its size
(cf. Table~\ref{tab:results}).


Values of $\delta$ in the range 2 to 5 involve very
modest synchrotron luminosities in the jet frame, equipartition
magnetic field strengths of order 1 mG, and a total energy stored in
particles and fields which can be supplied by a low drain on the jet
kinetic energy, $\approx10^{40}$erg~s$^{-1}$, a value $\leq$1\% of the
jet power estimated by Young, Wilson, \& Mundell (2002).  With these
parameters, the effective energy density of the CMB as seen by the jet
is less than the synchrotron energy density for most choices of
allowed $\Gamma$, and the combined photon energy density $\leq$0.2 of
that in the magnetic field for all $\delta\geq$2.  Thus the bulk of
the energy loss is via the synchrotron, not the IC channel.  The model
predicts that the optical emission will decay with a
timescale of order ten times longer than that we observe for the
X-rays since $\tau_{\frac{1}{2}}\propto\frac{1}{\gamma}$.  For a
straight jet, an angle to the line of sight of $\theta=10^{\circ}$
implies a total jet length of just under 10 kpc.

These estimates are based on decay timescales and our results would
change if the actual decay time is shorter than our current estimate,
in which case we would choose a smaller value of $\delta$; if the
magnetic field is substantially less than the equipartition value and
thus $\tau_{\frac{1}{2}}$ would be larger than calculated because of
the weaker field and we would again choose a smaller $\delta$; or if
the source size is substantially less than the light travel time which
would mean that the field is larger than calculated, halflives would
be shorter, and we would select a larger $\delta$.


\section{Conclusions}

We find that a synchrotron model with a beaming factor of $\delta$=2-5
gives a good fit to our observations for HST-1 and is compatible with
radio and optical interpretations.  It remains to be seen if similar
models can be constructed for powerful jets such as that in 3C~273.
Many of the gross features such as the progressions of X-ray emission
being stronger close to the core, whereas the radio intensity
increases moving out along the jet; and the slight shifts in
brightness (radio brighter downstream of the X-ray peak for some
knots) are the same in M87 and 3C273.  It is also worth noting that if
M87 were further away or observed with poor resolution, the
variability which we now know arises from a knot in the jet might well
be ascribed to conditions very close to the black hole (as in blazar
variability studies).  We shall continue to monitor the jet of M87
with Chandra between 2002 November and 2003 July with 8 observations
and the jet will also be monitored at similar intervals with the HST.
These data should provide refined estimates of variability timescales,
spectral evolution for the core and HST-1, and variability in the
other knots.




\acknowledgments

We thank the VLA and HST staffs for accommodating our ad hoc requests on
short notice.  The National Radio Astronomy Observatory is operated by
Associated Universities, Inc., under contract with the National
Science Foundation.  We thank O. Stohlman for assistance in the data
reduction and the referee for intelligent suggestions for improvement.
Work at SAO was supported by NASA contract NAS8-39073 and grant
GO2-3144X, and at U. Maryland by NASA grants NAG81027 and NAG81755.

\end{document}